\documentclass[11pt,a4paper]{article}
\usepackage{authblk}
\usepackage{natbib}
\usepackage[hyperref]{acl2020}
\usepackage{times}
\usepackage{latexsym}
\usepackage{booktabs}
\usepackage{multirow}
\usepackage{graphicx}
\usepackage{enumitem}
\usepackage{hyperref}
\usepackage{CJKutf8}
\usepackage{float}
\usepackage{subfigure}
\usepackage{array}
\usepackage{stfloats}
\usepackage{flushend}

\let\oldFootnote\footnote
\newcommand\nextToken\relax

\renewcommand\footnote[1]{%
    \oldFootnote{#1}\futurelet\nextToken\isFootnote}

\newcommand\isFootnote{%
    \ifx\footnote\nextToken\textsuperscript{,}\fi}

\aclfinalcopy 

\makeatletter
\def\UrlAlphabet{%
      \do\a\do\b\do\c\do\d\do\e\do\f\do\g\do\h\do\i\do\j%
      \do\k\do\l\do\m\do\n\do\o\do\p\do\q\do\r\do\s\do\t%
      \do\u\do\v\do\w\do\x\do\y\do\z\do\A\do\B\do\C\do\D%
      \do\E\do\F\do\G\do\H\do\I\do\J\do\K\do\L\do\M\do\N%
      \do\O\do\P\do\Q\do\R\do\S\do\T\do\U\do\V\do\W\do\X%
      \do\Y\do\Z}
\def\UrlDigits{\do\1\do\2\do\3\do\4\do\5\do\6\do\7\do\8\do\9\do\0}
\g@addto@macro{\UrlBreaks}{\UrlOrds}
\g@addto@macro{\UrlBreaks}{\UrlAlphabet}
\g@addto@macro{\UrlBreaks}{\UrlDigits}
\makeatother

\title{NAIST COVID: Multilingual COVID-19 Twitter and Weibo Dataset}

\author{\textbf{Zhiwei Gao}}
\author{\textbf{Shuntaro Yada}}
\author{\textbf{Shoko Wakamiya}}
\author{\textbf{Eiji Aramaki}}
\affil{Nara Institute of Science and Technology \authorcr  \texttt{\{gao.zhiwei.fw1,s-yada,wakamiya,aramaki\}@is.naist.jp}}

\date{}

\begin{document}

\maketitle
\begin{abstract}
Since the outbreak of coronavirus disease 2019 (COVID-19) in the late 2019, it has affected over 200 countries and billions of people worldwide. This has affected the social life of people owing to enforcements, such as ``social distancing'' and ``stay at home.'' This has resulted in an increasing interaction through social media. Given that social media can bring us valuable information about COVID-19 at a global scale, it is important to share the data and encourage social media studies against COVID-19 or other infectious diseases. 
Therefore, we have released a multilingual dataset of social media posts related to COVID-19, consisting of microblogs in English and Japanese from \textit{Twitter} and those in Chinese from \textit{Weibo}.
The data cover microblogs from January 20, 2020, to March 24, 2020.
This paper also provides a quantitative as well as qualitative analysis of these datasets by creating daily word clouds as an example of text-mining analysis. The dataset is now available on Github.\footnote{ \url{https://github.com/sociocom/covid19_dataset}} This dataset can be analyzed in a multitude of ways and is expected to help in efficient communication of precautions related to COVID-19.
\end{abstract}

\section{Introduction}
The outbreak of the coronavirus disease 2019 (COVID-19) was observed at the end of 2019 in Wuhan, Hubei Province, China. Since January 2020, it has rapidly spread worldwide. On March 11, 2020, the World Health Organization (WHO) announced that COVID-19 can be characterized as a pandemic. The virus causing COVID-19, severe acute respiratory syndrome coronavirus-2 (SARS-CoV-2), has infected more than 1.2 million people worldwide, and 60,000 people have lost their lives.\footnote{\url{https://google.com/covid19-map/}} WHO highly recommends maintaining ``social distancing'' measures, and several countries with severe epidemics are further requesting citizens to stay home.

In this scenario, online social media, such as \textit{Twitter}, \textit{Weibo}, and \textit{Instagram}, are playing an important role in sharing information and perception about COVID-19.
Social media is recognized as one of the valuable resource of data that can lead to prediction of various phenomena related to an event. For example, \citet{5604088} showed that microblog data facilitated better public-health surveillance, such as the prediction of the number of patients suffering from influenza.

To encourage and support the social media studies on COVID-19, it is crucial to make relevant datasets available to the public.
Here, we publish a multilingual dataset that contains over 20 million microblogs related to COVID-19 in English, Japanese, and Chinese from \textit{Twitter} and \textit{Weibo} since January 20, 2020, until March 24, 2020. 

\citet{chen2020covid19} and \citet{lopez2020understanding} have already released multilingual datasets collected from \textit{Twitter}. Given that China is the very first country to have faced a COVID-19 outbreak, we further collected microblogs about COVID-19 from \textit{Weibo}, one of the most popular social media in China similar to \textit{Twitter}.

The remainder of the paper is organized as described follows. 
In Section~\ref{sec:collect}, we elaborate on the method of data collection.
In Section~\ref{sec:analysis}, we provide a quantitative analysis of the dataset, such as the character count per microblog and the microblog count per day. 
In Section~\ref{sec:analysis2}, we present the daily word cloud images created from microblogs of each language as an example of text-mining analysis. 
Finally, in Section~\ref{sec:conclusion}, we present the conclusion with our future work. 

\section{Data Collection} \label{sec:collect}

\begin{table*}
\centering
\resizebox{\textwidth}{!}{%
\begin{tabular}{@{}cccc@{}}
\toprule
\multicolumn{1}{l}{} & \textbf{Phase~1} & \textbf{Phase~2} & \textbf{Phase~3} \\ \midrule
\multirow{3}{*}{English} & \multirow{3}{*}{Wuhan AND (pneumonia OR coronavirus)} & \multirow{3}{*}{Wuhan AND (pneumonia OR coronavirus OR (COVID AND 19))} & (Wuhan AND pneumonia) OR\\
 &  &  & coronavirus OR\\
 &  &  & (COVID AND 19) \\ \midrule
\multirow{3}{*}{Japanese} & \multirow{3}{*}{\begin{CJK*}{UTF8}{ipxm}武漢~AND (肺炎~OR~コロナ)\end{CJK*}} & \multirow{3}{*}{\begin{CJK*}{UTF8}{ipxm}武漢~AND (肺炎~OR コロナ~OR (COVID AND 19))\end{CJK*}} & \begin{CJK*}{UTF8}{ipxm}(武漢~AND~肺炎)\end{CJK*} OR\\
 &  &  & \begin{CJK*}{UTF8}{ipxm}コロナ\end{CJK*} OR\\
 &  &  & \begin{CJK*}{UTF8}{ipxm}(COVID AND 19)\end{CJK*} \\ \midrule
\multirow{3}{*}{Chinese} & \multirow{3}{*}{\begin{CJK*}{UTF8}{gbsn}武汉~AND (肺炎~OR~冠状病毒)\end{CJK*}} & \multirow{3}{*}{\begin{CJK*}{UTF8}{gbsn}武汉~AND (肺炎~OR~冠状病毒~OR~新冠肺炎)\end{CJK*}} & \begin{CJK*}{UTF8}{gbsn}(武汉~AND~肺炎)\end{CJK*} OR \\
 &  &  & \begin{CJK*}{UTF8}{gbsn}冠状病毒\end{CJK*} OR \\
 &  &  & \begin{CJK*}{UTF8}{gbsn}新冠肺炎\end{CJK*} \\ \bottomrule
\end{tabular}%
}
\caption{Keywords used for collecting English, Japanese, and Chinese microbloogs in each phase. AND OR denote search operators.}
\label{tab:keywords}
\end{table*}

To collect the microblogs related to COVID-19, we adopted keyword-based search.
For English and Japanese, we collected microblogs related to COVID-19 from \textit{Twitter}, while we obtained Chinese microblogs from \textit{Weibo}. We employed \textit{Twitter} Search API\footnote{\url{https://developer.twitter.com/en/docs/tweets/search/overview/standard}} for tweets; a web crawler was applied to retrieve \textit{Weibo} posts.

\subsection{Keywords}
We developed three sets of query keywords as shown in Table~\ref{tab:keywords} according to the stages of COVID-19 spread.
Corresponding to these sets, our dataset can be divided into three phases: 
\begin{description}
    \item[Phase~1] (January 20 to February 23, 2020):\\ 
        In combination with the term ``Wuhan,'' we used the keywords ``pneumonia'' and ``coronavirus'' in English and their translations in Japanese and Chinese. 
        We included the Chinese city name ``Wuhan'' as the primary keyword, because Wuhan (\begin{CJK*}{UTF8}{ipxm}``武漢''\end{CJK*} in Japanese and \begin{CJK*}{UTF8}{gbsn}``武汉''\end{CJK*} in Chinese) observed the earliest outbreak with the maximum number of confirmed cases. 
        Note that in the said period, the official disease name ``COVID-19'' was yet to be defined.
    \item[Phase~2] (February 24 to 29, 2020):\\ 
        WHO assigned the official name ``COVID-19'' on February 11. 
        We added it to the keywords in combination with ``Wuhan,'' 
        although this resulted in a smaller number of retrieval because all the microblogs included ``Wuhan.''
    \item[Phase~3] (March 1–24, 2020):\\
        To obtain more data, we relaxed search conditions by querying each set of keywords separately.
\end{description}

\begin{table}
\centering
\resizebox{0.4\textwidth}{!}{%
\begin{tabular}{@{}lrrr@{}}
\toprule
 & \multicolumn{1}{c}{\textbf{Phase~1}} & \multicolumn{1}{c}{\textbf{Phase~2}} & \multicolumn{1}{c}{\textbf{Phase~3}} \\ \midrule
English & 247,350 & 41,647 & 15,961,041 \\
Japanese & 233,065 & 4,0953 & 9,227,848 \\
Chinese & 84,647 & 18,750 & 70,472 \\ \bottomrule
\end{tabular}%
}
\caption{Number of microblogs in each language during different phases.}
\label{tab:count}
\end{table}

\subsection{Data Size}

As shown in Table~\ref{tab:count}, we have collected over 16 million microblogs in English, 9 million in Japanese, and 180 thousand in Chinese during January 20 to March 24, 2020.
To collect \textit{Twitter} and \textit{Weibo} posts, we have adopted a uniform daily timing to collect microblogs from 0:00 to 23:59 (JST) of the previous day.
To ensure the uniqueness of the data, for \textit{Twitter}, we filtered out all retweets by adding the ``-filter:retweets'' operator;
for \textit{Weibo}, we searched for ``original microblogs'' only.
Note that we have collected smaller amounts of the data from \textit{Weibo} than \textit{Twitter} because anti-crawling mechanism in \textit{Weibo} limits our web crawler to access only the first 50 pages of the search content. 

\subsection{Dataset Accessibility}

We released the first version of the dataset on Github at \url{https://github.com/sociocom/covid19_dataset}. Following the terms of service of \textit{Twitter} and \textit{Weibo}, we mainly published microblog IDs, instead of exposing original text and metadata. The dataset consists of the lists of microblog IDs with two fields of metadata: their timestamps and the query keywords mentioned in the microblogs among our search queries. This helps make subsets suitable for subsequent applications and tasks.
Since a \textit{Weibo}'s microblog is uniquely determined by the combination of user ID and microblog ID, we share the corresponding user ID and microblog ID for each microblog in the form of ``user ID/microblog ID.''

\section{Quantitative Analysis} \label{sec:analysis}

We provide basic statistics of our dataset in terms of its quantitative volume.
First, we show the number of characters in microblogs.
Next, we plot the number of microblogs per time series.

\subsection{Character Count}

While microblogs contain multimodal data (e.g., images and movies), their core content is text.
We report the number of characters to quantify the total amount of our dataset.
Table~\ref{tab:avg-std} shows the sum, mean, and standard deviation of the number of characters for each language in our dataset.
We removed URLs and punctuations from each microblog to expose the amount of characters that constituted the essential content.

\begin{table}[t]
\centering
\resizebox{0.48\textwidth}{!}{%
\begin{tabular}{@{}lrrr@{}}
\toprule
\multicolumn{1}{c}{\textbf{Language}} & \multicolumn{1}{c}{\textbf{Sum}} & \multicolumn{1}{c}{\textbf{Mean}} & \multicolumn{1}{c}{\textbf{Standard deviation}} \\ \midrule
English & 2,268,395,730 & 139.59 & 75.90 \\
Japanese & 626,113,353 & 65.89 & 38.19 \\
Chinese & 25,115,113 & 144.45 & 169.69 \\ \bottomrule
\end{tabular}%
}
\caption{Statistics of characters for each language in our dataset.}
\label{tab:avg-std}
\end{table}

\subsection{Daily Microblog Count} \label{sec:daily count}

\begin{figure*}[!ht]
    \centering
    \subfigure[The number of English microblogs and the daily confirmed cases in major English-speaking countries in January and February.]{
        \includegraphics[width=2.8in]{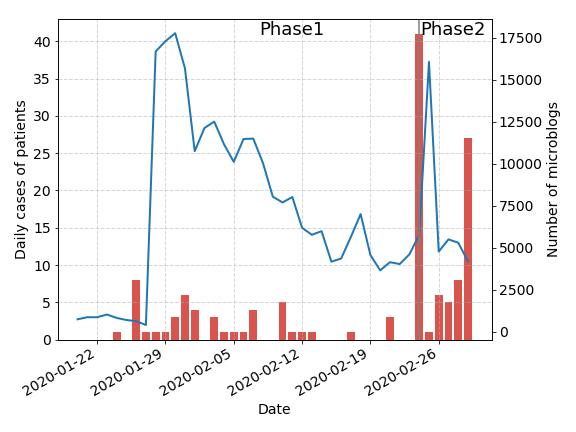} 
        \label{fig:en-jan}
    }
    \subfigure[The number of English microblogs and the daily confirmed cases in major English-speaking countries in March.]{
        \includegraphics[width=2.8in]{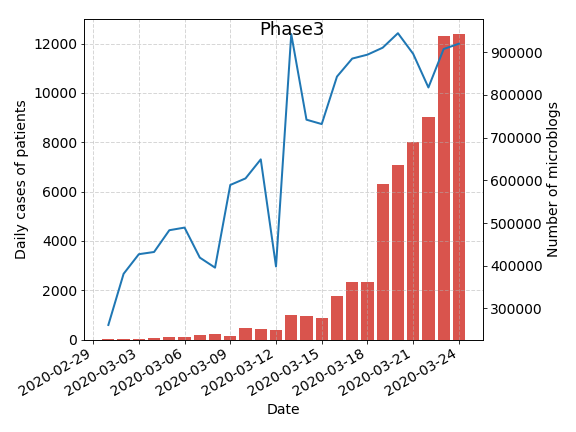}
        \label{fig:en-mar}
    }
    \subfigure[The number of Japanese microblogs and the daily confirmed cases in Japan in January and February.]{
        \includegraphics[width=2.8in]{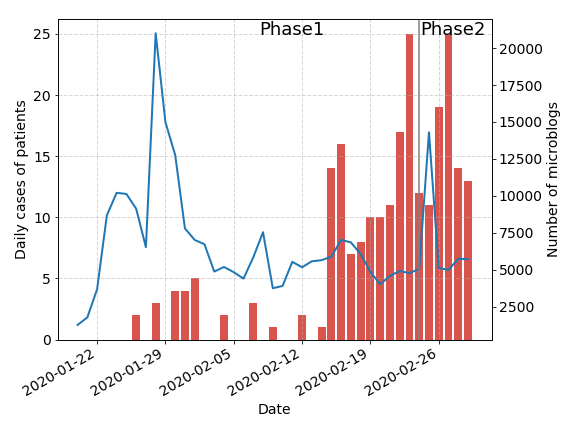}
        \label{fig:ja-jan}
    }
    \subfigure[The number of Japanese microblogs and the daily confirmed cases in Japan in March.]{
        \includegraphics[width=2.8in]{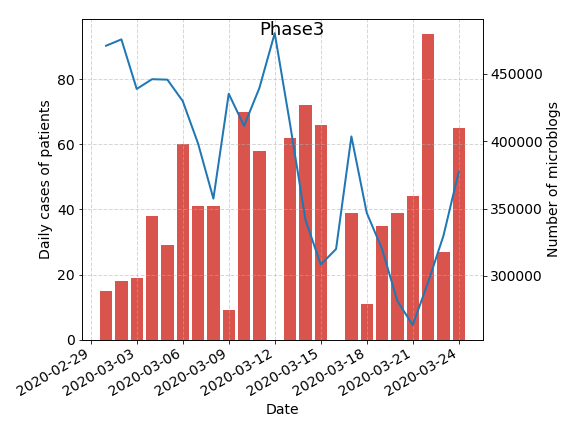}
        \label{fig:ja-mar}
    }
    \subfigure[The number of Chinese microblogs and the daily confirmed cases in China in January and February.]{
        \includegraphics[width=2.8in]{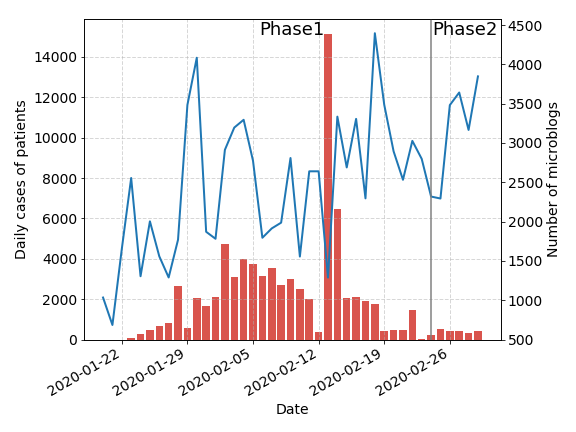}
        \label{fig:zh-jan}
    }
    \subfigure[The number of Chinese microblogs and the daily confirmed cases in China in March.]{
        \includegraphics[width=2.8in]{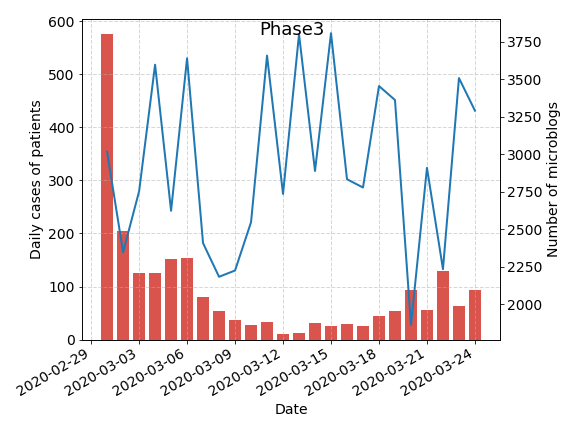}
        \label{fig:zh-mar}
    }
    \caption{Analysis of the day-by-day count of microblogs for each language. Solid line represents the timeline of the number of microblogs and red bar represents the number of the daily confirmed cases of COVID-19-positive patients.}
    \label{fig:statistics} %
\end{figure*}

Figure~\ref{fig:statistics} portrays the daily count of microblogs in each language, combined with the number of confirmed cases of COVID-19 patients every day, which is obtained from DataHub.io\footnote{\url{https://datahub.io/core/covid-19}}.
Figure~\ref{fig:en-jan} is the plot of English microblogs and the confirmed cases in four major English-speaking countries (i.e., Australia, Canada, the United Kingdom, and the United States) during Phases~1 and 2;
Figure~\ref{fig:en-mar} shows that in Phase~3.
Figures~\ref{fig:ja-jan} and~\ref{fig:zh-jan} are the Japanese and Chinese versions of the same plots for Phases~1 and 2, whereas Figures~\ref{fig:ja-mar} and~\ref{fig:zh-mar} display the plots of Phase~3.


In Figure~\ref{fig:en-jan}, a sudden and dramatic increase in the number of English microblogs can be observed on January 28, 2020.
According to the news, that particular day saw a discussion on the death toll in mainland China reaching 100.\footnote{January 28, 2020; CNN, \url{https://cnn.it/3a1FFm8}} On the same day, Japan also observed a sharp rise in the relevant microblogs, as shown in Figure~\ref{fig:ja-jan}. This was a result of many users tweeting extensively about the three newly confirmed cases in Japan, which included people who had not been to Wuhan.\footnote{January 28, 2020; Japan Times, \url{https://bit.ly/3aFPqaE}}

Subsequently, there was a substantial increase in the English microblogs on February 25, 2020, as shown in Figure~\ref{fig:en-jan}. On that day, there were reports that ``Trump privately vents over his team's response to coronavirus -- even though he says that the virus is under control,''\footnote{February 25, 2020; CNN, \url{https://cnn.it/39VVbjg}} leading to many microblogs against Trump on \textit{Twitter}.

In March, as Figure~\ref{fig:en-mar} shows, the number of microblogs in major English-speaking countries showed an upward trend as the number of the confirmed cases increased, and the largest number of microblogs exceeded 9 million a day. 
Meanwhile, in Japan, the number of daily confirmed cases was relatively small as shown in Figure~\ref{fig:ja-mar}. Therefore, we assumed that Japanese \textit{Twitter} users are not as interested in COVID-19 as in the major English-speaking countries. In particular, there was a decline in the number of microblogs from March 12 to March 15, 2020. March 12, 2020, was the Olympic flame lighting ceremony and the torch relay for the Tokyo 2020 Olympics.\footnote{March 12, 2020; BBC, \url{https://bbc.in/3emD6OK}} Therefore, we speculate that this sudden decrease was caused by a shift in attention from COVID-19 to the torch relay for many Japanese users.


With regard to the Chinese microblogs, the trends of the numbers are shown in Figures~\ref{fig:zh-jan} and~\ref{fig:zh-mar}. These do not fully reflect the quantitative trends of the confirmed cases owing to the limited amount of the microblogs we could collect on a daily basis.

\section{Qualitative Analysis} \label{sec:analysis2}
\begin{figure*}[tp]
    \centering
    \subfigure[Word cloud of English microblogs on February 8, 2020.]{
        \includegraphics[width=3in]{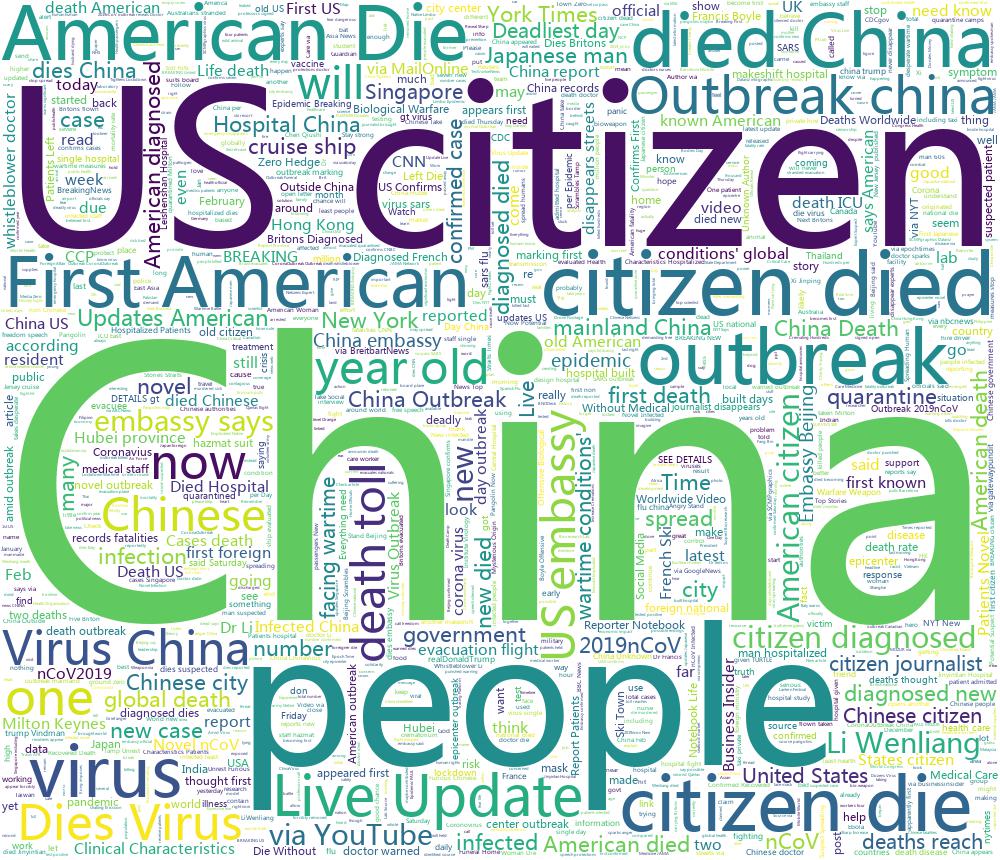}
        \label{fig:en0208}
    }
    \subfigure[Word cloud of English microblogs on March 16, 2020.]{
        \includegraphics[width=3in]{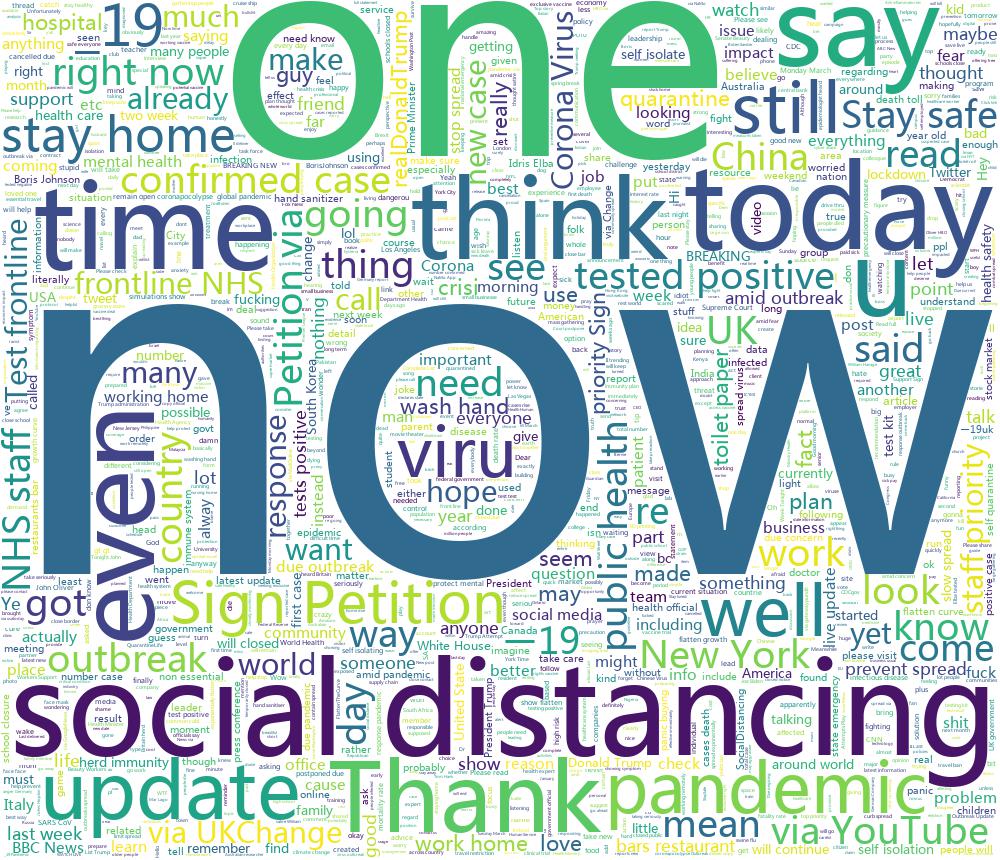}
        \label{fig:en0316}
    }
    \subfigure[Word cloud of Japanese microblogs on January 28, 2020.]{
        \includegraphics[width=3in]{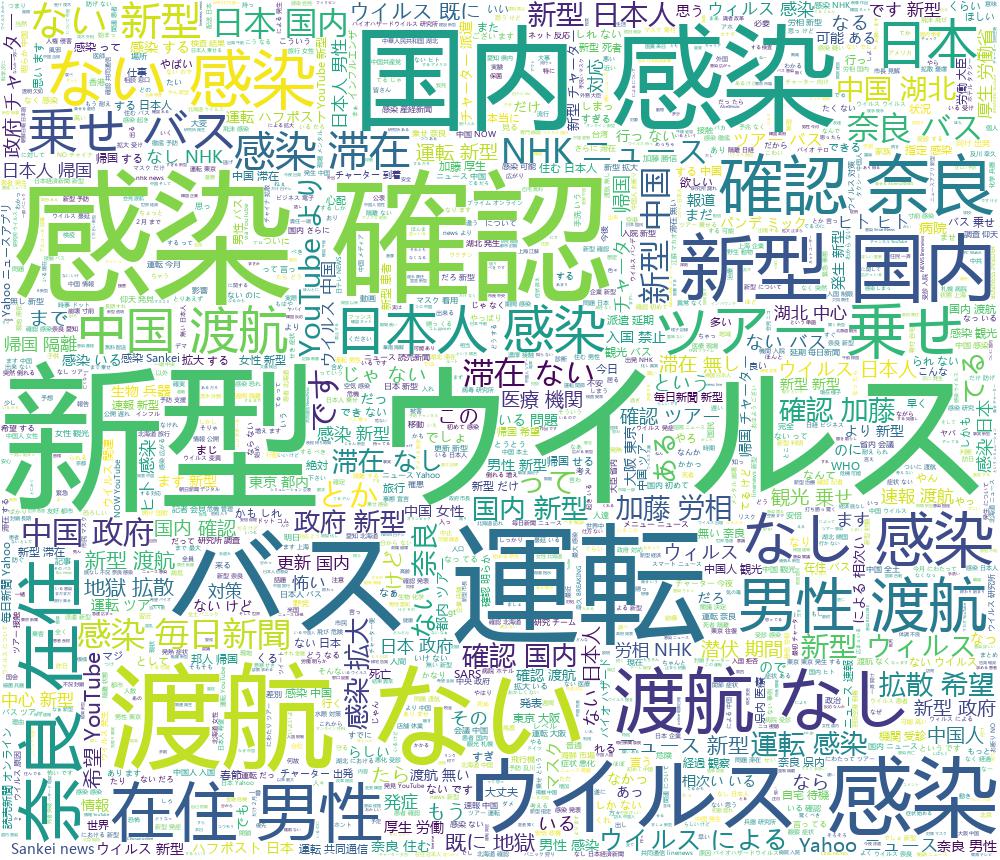}
        \label{fig:ja0128}
    }
    \subfigure[Word cloud of Japanese microblogs on March 24, 2020.]{
        \includegraphics[width=3in]{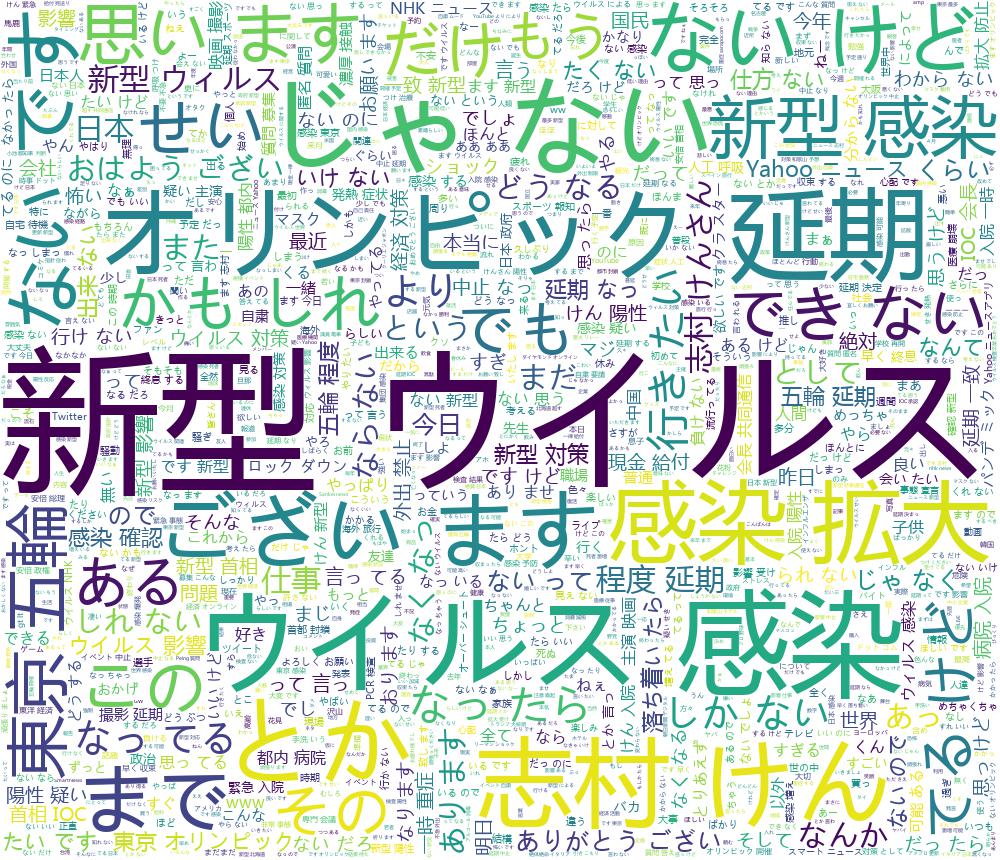}
        \label{fig:ja0324}
    }
    \subfigure[Word cloud of Chinese microblogs on January 20, 2020.]{
        \includegraphics[width=3in]{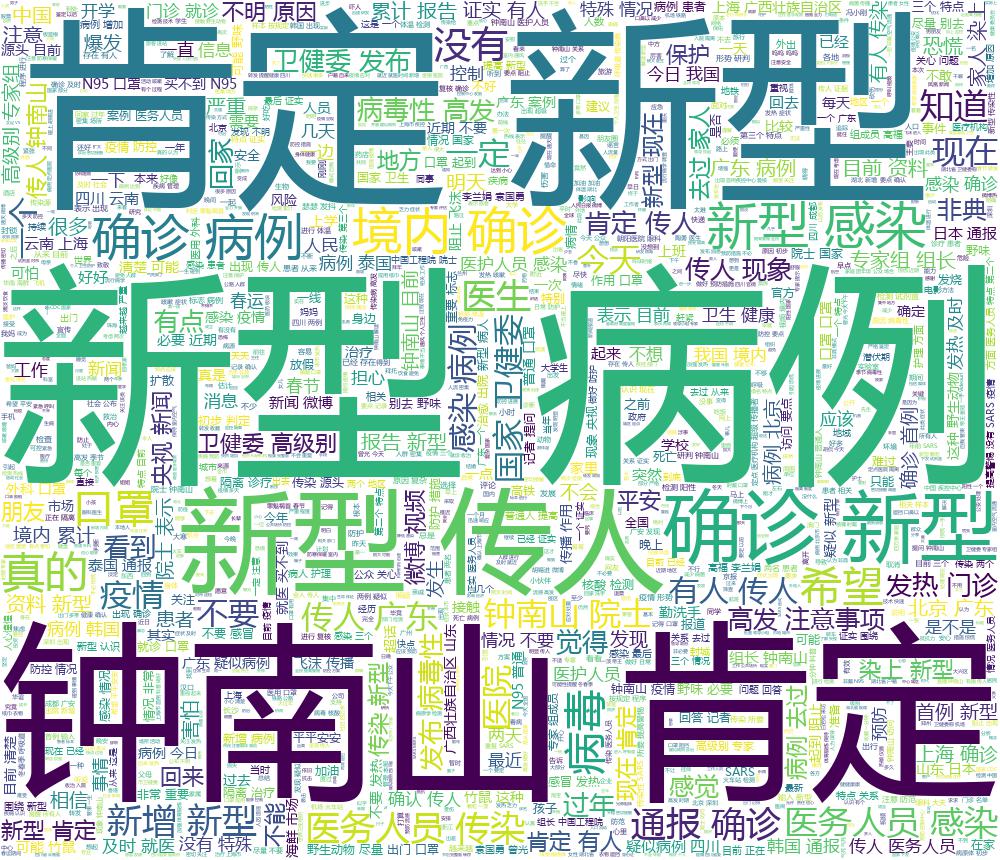}
        \label{fig:zh0120}
    }
    \subfigure[Word cloud of Chinese microblogs on March 10, 2020.]{
        \includegraphics[width=3in]{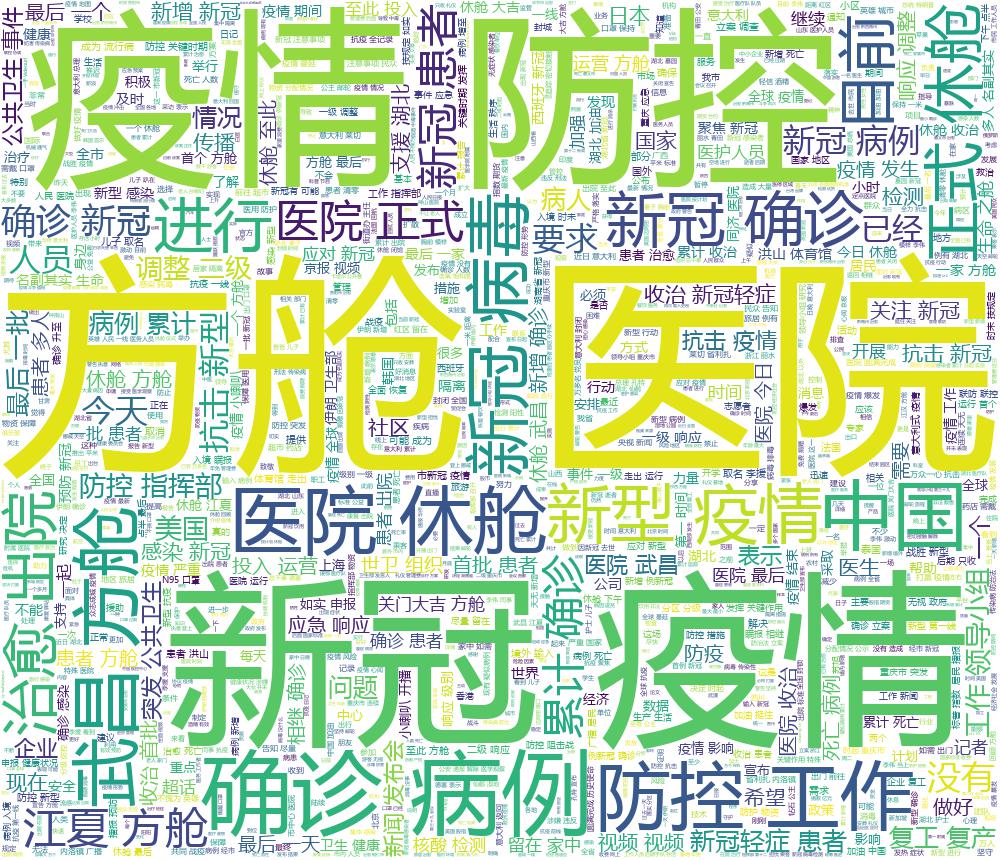}
        \label{fig:zh0310}
    }
    \caption{Daily word cloud images for each language.}
    \label{fig:wordclouds} %
\end{figure*}

In addition to the quantitative analysis, we show an example of qualitative analysis based on our dataset.
As an initial attempt, we adopted a \textit{word cloud}, which is ``an electronic image that shows words used in a particular piece of electronic texts or series of texts.''\footnote{\url{https://dictionary.cambridge.org/dictionary/english/word-cloud}}
In word clouds, term frequency for each word in a corpus is proportional to its font size, which enables us to grasp the topics of the corpus visually.
Daily word cloud images of our dataset for each language are available at \url{https://aoi.naist.jp/2020-covid/wordcloud}.
Henceforth, we provide brief interpretations of these word clouds to demonstrate a possible text-mining approach that can be applied to our dataset in Figure~\ref{fig:wordclouds}.

Note that we removed stop words followed by tokenization in our word clouds. 
For the Chinese and Japanese tokenization, we used Jieba\footnote{\url{https://github.com/fxsjy/jieba}} and Mecab\footnote{\url{https://taku910.github.io/mecab}}, respectively. 
We also filtered out the search keywords in each microblog to reduce the disturbance of these keywords in the image.

\subsection{English Word Cloud}
    A US citizen who lived in Wuhan passed away because of COVID-19 in Wuhan on February 8, 2020.\footnote{February 8, 2020; CNBC, \url{https://cnb.cx/2R4uYZ1}}
    This was the first casualty of a US citizen.
    The word cloud of this day, shown in Figure~\ref{fig:en0208}, contains the related words, e.g., ``American,'' ``US,'' ``citizen,'' and ``die.''
    
    Figure~\ref{fig:en0316} is the word cloud on March 16, 2020, in which ``social distancing,'' an important phrase to fight against the epidemic, appears notably.
    We can also notice that another socially important phrase ``stay home'' has an increased in size in our word cloud series from March 20, 2020.

\subsection{Japanese Word Cloud}
    The first local transmission of COVID-19 inside Japan was reported on January 28, 2020, as described in Section~\ref{sec:daily count}.
    Figure~\ref{fig:ja0128} shows the word cloud on that day. It reflects the fact that the infected patient lived in Nara prefecture and drove a sightseeing-tour bus that carried travelers from Wuhan. We can observe the relevant keywords, such as 
    \begin{CJK*}{UTF8}{ipxm}
    ``奈良~(Nara),'' ``バス~(bus),'' and ``運転~(drive).''
    \end{CJK*}
    
    On March 24, 2020, Japan and International Olympic Committee (IOC) officially agreed to postpone the planned 2020 Tokyo Olympics until 2021.\footnote{March 24, 2020; The Washington Post, \url{https://wapo.st/2UYXEnG}}
    A notable change in Japanese word cloud series can be found as the novel appearance of the words 
    \begin{CJK*}{UTF8}{ipxm}
    ``オリンピック~(Olympics)'' and ``延期~(postponing)'' in that day's figure (i.e., Figure~\ref{fig:ja0324}).
    \end{CJK*}
    
    We can also notice that a YouTube video became viral in Japanese \textit{Twitter} from around January 29 to February 6, 2020, by observing the corresponding word clouds. The video was originally made by a Wuhan citizen and subtitled in Japanese later by another YouTuber,\footnote{January 29, 2020; YouTube, \url{https://youtu.be/Mcfn5Eh5OVE}} which tells the situation of Wuhan in lockdown.
    In addition to the word ``YouTube,'' the corresponding word clouds contain the tokens of the video title, i.e., 
    \begin{CJK*}{UTF8}{ipxm}
    ``震源~(hypocenter),'' ``動画~(video),'' and ``和訳~(Japanese translation).''
    \end{CJK*}

\subsection{Chinese Word Cloud}
     Figure~\ref{fig:zh0120} shows the word cloud on January 20, 2020, and also shows that the term \begin{CJK*}{UTF8}{gbsn}
     ``钟南山~(Zhong nanshan)''
     \end{CJK*}
     has a larger weight. It was on January 20 that Dr.~Zhong indicated the existence of human-to-human transmission of COVID-19\footnote{January 20, 2020; The New York Times, \url{https://nyti.ms/3bT7r5m}} that triggered extensive discussion on \textit{Weibo}. \par
    Figure~\ref{fig:zh0310} shows the word cloud on March 10, 2020 and the word \begin{CJK*}{UTF8}{gbsn}
    ``方舱医院~(mobile cabin hospital)''
    \end{CJK*}
    was more conspicuous. According to China's National Health Commission, all of Wuhan's mobile cabin hospitals were closed on March 10.\footnote{March 10, 2020; Xinhua News, \url{https://bit.ly/2JG28u6}} The mobile cabin hospitals, which were instrumental in preventing the spread of the epidemic, also had attracted much attention.



\section{Conclusion}  \label{sec:conclusion}

We published a multilingual dataset of microblogs related to COVID-19 collected by relevant query keywords at \url{https://github.com/sociocom/covid19_dataset}.
The dataset covered English and Japanese tweets from \textit{Twitter} and Chinese posts from \textit{Weibo}.
The present version of the dataset (\today) encompassed microblogs from January 20 to March 24, 2020.

We then showed one of the possible utilization of our dataset through the daily microblog count analysis as an example of the quantitative analyses and the word cloud-based analysis as an example of the qualitative analyses. The results of the analyses are summarized as follows. 
For China, which is the first country to have faced a full-blown outbreak of COVID-19, we can observe from social media that people took the situation and prevention seriously. As the number of confirmed cases in China decreased, the trend in social media shifted toward the concern for the global situation. 
In the UK and the US, the main English-speaking countries, initially, there was less social media interests owing to fewer confirmed cases. The subsequent outbreaks sprung the discussion about COVID-19 on social media, including the promotion of precautionary measures and recommendations to keep “social distancing” measures. Meanwhile, Japan showed relatively sluggish growth. However, on March 24, 2020, the announcement of the postponement of the 2020 Olympic Games in Tokyo along with a relatively rapid growth of confirmed cases was reflected in the increased social media activity. This was accompanied by microblogs expressing concerns about the epidemic and dissatisfaction with government measures.\par

We believe that this dataset can be analyzed further in many ways, such as sentiment-based analysis\footnote{\url{https://usc-melady.github.io/COVID-19-Tweet-Analysis/}}, comparison with web search queries, moving logs\footnote{\url{https://www.google.com/covid19/mobility/}}\footnote{\url{https://dataforgood.fb.com/tools/disease-prevention-maps}}, etc.
Various combinations of data can enable deeper analyses of social media communication. 
Furthermore, our dataset would contribute to extract useful clinical information from social media and render hints about efficient broadcasting of the clinical information.
We continue to collect the microblog data while keeping the repository up-to-date. \par

\section*{Acknowledgments}
This study was supported in part by JSPS KAKENHI Grant Number JP19K20279 and Health and Labor Sciences Research Grant Number H30-shinkougyousei-shitei-004.

\bibliography{references}
\bibliographystyle{acl_natbib}

\end{document}